\documentclass[%
reprint,
showpacs,
amsmath,amssymb,aps,
]{revtex4-1}

\usepackage{graphicx}
\usepackage{dcolumn}
\usepackage{bm}
\usepackage{color}

\newcommand{\md}{\mathrm d}
\newcommand{\etag}{\eta_{\rm g}}
\newcommand{\mg}{\mu_{\rm g}}
\newcommand{\mq}{\mu_{\rm q}}
\newcommand{\mgs}{\mu_{\rm g}^2}
\newcommand{\pg}{\phi_{\rm g}}
\newcommand{\pq}{\phi_{\rm q}}
\newcommand{\eg}{E_{\rm g}}
\newcommand{\Mg}{M_{\rm g}}

\begin{document}


\title{Effective model based on QCD with gluon condensate}%

\author{Hiroaki Kohyama}
\affiliation{Department of Physics,
National Taiwan University, Taipei 10617, Taiwan}

\date{\today}

\begin{abstract}
We construct the effective model based on QCD with gluon
condensate. Under the assumption that the gluons are condensed with
the sharp momentum peak in the momentum space, we formulate the
effective field theory incorporating both the gluon condensate and the
chiral condensate, then study the phase transition on temperature and
chemical potential plane with respect to two
condensates. We find that the condensates decrease with increasing
temperature, which is reliable tendency on the condensate
being consistent with the argument of the asymptotic behavior.
\end{abstract}

\pacs{11.10.Wx, 12.38.-t, 12.40.-y}
\maketitle

\section{\label{sec:intro}%
Introduction}
The first principle theory for quarks and gluons is quantum
chromodynamics (QCD), whose goal is to explain all the phenomena
relating to the strong interaction. However, it is known to be difficult
since usual perturbative technologies in the quantum theory calculations
hardly works for low energy physics in which nonperturbative effect
dominates the system then complicates the analyses. One of
characteristic features is non-trivial gluon and chiral condensates
in the nonperturbatively correlated system due to the strong
interaction. It is challenging to derive the condensates relying on
the perturbative quantum field analyses, then one usually has to use
some effective models of QCD for the sake of describing the low
energy physics, such as properties of hadrons.

There has been a lot of investigations on low energy phenomena
based on various effective theories treating the gluon and
chiral condensates.
The instanton includes the gluon condensate though the solution
to the classical field equation~\cite{Belavin:1975fg}, and various
applications have been done~%
\cite{Callan:1977gz, Vainshtein:1981wh, Schafer:1996wv,
Shuryak:1997vd}.
QCD sum rule is formulated by the expansion of the correlation
functions in the vacuum condensates~\cite{Shifman:1978by},
which is as well widely used for many applications%
~\cite{Reinders:1984sr, Cohen:1994wm, Shifman:1998rb}.
The Schwinger-Dyson equations (SDE) are the equations for
the exact correlation functions of fields~\cite{Dyson:1949ha},
and there has been devoted to a lot of analyses%
~\cite{Roberts:1994dr, Roberts:2000aa, Maris:2003vk,
Fischer:2006ub}.
The Nambu--Jona-Lasinio (NJL) model~\cite{Nambu:1961tp}
is constructed using the analogy with the BCS theory for
superconductivity~\cite{Bardeen:1957mv}; the model studies the
hadron properties based on the back ground chiral condensate%
~\cite{Vogl:1991qt, Klevansky:1992qe, Hatsuda:1994pi,
Huang:2004ik, Buballa:2005rept}.
An extension of the NJL model has been made
in~\cite{Fukushima:2003fw} where the Polyakov-loop contribution
is incorporated into the chiral dynamics, and extensive analyses on
both the chiral and gluon loop condensates have been performed in
the Polyakov loop NJL (PNJL) model~\cite{Fukushima:2008wg}.
Thus the issues on the gluon and chiral condensates are
essential in the study of hadron physics, there we believe these
two condensates are intimately related to each other.

In this paper, we construct the effective model based on QCD
by postulating that gluons form the condensation whose momenta
have the sharp peak at the QCD scale~\cite{Kohyama:2016dhd}.
The speculation is inspired by the experiments 
of the momentum in the Bose Einstein condensate (BEC) state where
the clear peak in the momentum distribution is actually observed.
We then consider the analogy between the BEC state and the gluon
condensed state as done in a chiral effective model motivated from
the BCS analogy.  To be more concrete, we are going to formulate
the gap equation for the gluon condensate from the QCD under the
above mentioned assumption with gluon condensate.
It may be important to remark that the solution of the gap equation
for the gluon condensate at high temperature behaves badly, since
many bosons can unlimitedly occupy the same point due to
the statistical property of bosonic fields to cause the catastrophic
situation in the gap equation. Therefore the special attention should
be paid when one treats the solution at high temperature near the phase
transition. However, although the high temperature results do not
behave nicely, meaning that the plots deviate from the expected
ones from the BEC experimental data, we find that the tendency of
the gluon condensate shows physically reliable feature as will be
presented later. We then proceed the analyses based
on the gluon condensate by using the gap equation with simple form. 
Here the purpose of this letter is twofold: first, to
construct the QCD motivated effective model based on the anzatz
of the gluon condensate and second, to see how the obtained
model describes the thermodynamic properties and how it affects to
the chiral condensate.

The paper is organized following. We formulate the gap equations
for the gluons, quarks, and coupled system in Sec. \ref{sec:qcd}. The
numerical solutions for the gap equations in the separate models
are shown in Sec. \ref{sec:separate}. Secs. \ref{sec:coupled} and
\ref{sec:chiral_phase} are devoted to the investigations on the
thermodynamic quantities and the phase structure in the coupled
model. The conclusions and some subsidiary calculations
are given in Sec. \ref{sec:conclusion} and Appendices.

\section{\label{sec:qcd}%
QCD with gluon condensate
}
Starting from the QCD Lagrangian, we are going to evaluate
the effective potential under the hypothetical situation of the
condensed gluons. Thereafter the gap equations for the gluon
and chiral condensates, and their coupled equation are derived.

\subsection{\label{subsec:g_gap}%
Gluon gap equation}
The Lagrangian for gluon is given by
\begin{align}
 & \mathcal{L}_{\rm g}
   =  \mathcal{L}_{\rm g}^0
    +\mathcal{L}_{\rm g}^3
    +\mathcal{L}_{\rm g}^4, \\
 & \mathcal{L}_{\rm g}^0
   = 
      -\frac{1}{4}(\partial_\mu A^a_\nu - \partial_\nu A^a_\mu)^2,\\
 & \mathcal{L}_{\rm g}^3
      = - gf^{abc}(\partial_\mu A_\nu^a) A^{\mu b}A^{\nu c}, \\
 & \mathcal{L}_{\rm g}^4
      = -\frac{1}{4} g^2
        (f^{eab}A^{a}_\mu A^{b}_\nu)(f^{ecd}A^{\mu c}A^{\nu d}),
\end{align}
where $A_\mu$ is the gluon field, $g$ is the coupling
constant for the strong interaction and $f^{abc}$ is the
structure constant with respect to the ${\rm SU}(3)$ color
system.

The effective potential can be evaluated through the relation,
$\Omega_{\rm g} \equiv -\ln {\mathcal Z}/V$ with the partition
function ${\mathcal Z}$
and the volume of the system $V$, whose explicit form becomes
\begin{align}
  {\Omega}_{\rm g}
  = -\frac{1}{V}
  \int \md^4 x \,\,  \langle \mathcal{L}_{\rm g} \rangle
  - \frac{i}{2V} \ln \det 
  \left\langle
     -\frac{\delta^2 \mathcal{L}_{\rm g}}{\delta A_\mu^a \delta A_\nu^b} 
  \right\rangle
  + \cdots
\label{eq:O_A}
\end{align}
where the bracket expresses the
expectation value of the quantity, $O_{\rm ex} \equiv \langle O \rangle$.

Before proceeding the calculation of the effective potential, we present
the main assumption on the gluon fields here. The characteristic
treatment of the gluon field is that we assume the fields are condensed
then have non-zero expectation value. The usual Feynman rule for the
two-point function on the gluon is the amplitude 
\begin{align}
  \bigl\langle
     A^a_{\mu} (x)  A^b_{\nu} (x)
  \bigr\rangle_{\rm p}
  = g_{\mu \nu} \delta^{ab}
    \int \frac{\md^4 p}{(2\pi)^4}
    \frac{-i}{p^2 + i\epsilon}.
\end{align}
calculated from the perturbative gluon propagator.
The amplitude badly diverges due to the loop integral, then we need to
perform the renormalization so that one gets finite physical predictions. 
Here, we postulate that the exact renormalized two-point function has
the following expectation value,
\begin{equation}
  \langle A_\mu^a(x) A_\nu^b(x) \rangle_{\rm ex}
  = g_{\mu\nu} \delta^{ab} \pg,
\label{eq:g_cond}
\end{equation}
which is made to be finite by virtue of the renormalization,
and we call the quantity $\pg$ as the gluon condensate.
Eq.(\ref{eq:g_cond}) is the key hypothesis employed in this paper.

Let us now carry out the evaluation of the effective potential
with the help of the above mentioned ingredients.
The non-zero contribution in the first term of Eq.~(\ref{eq:O_A})
stems from the four-point interaction~\cite{Kohyama:2016dhd},
\begin{equation}
 \langle \mathcal{L}_{\rm g} \rangle
 = -72 g^2 \pg^2.
\end{equation}
Note that $\langle \mathcal{L}_{\rm g}^3 \rangle =0$ due to the
antisymmetric property of the structure constant $f^{abc}$.
It is also straightforward to evaluate the second term in
Eq. (\ref{eq:O_A}), which reads
\begin{align}
  & \frac{-i}{2V} \ln \det
      \left\langle -
        \frac{\delta^2 {\mathcal L_{\rm g}}}{\delta A_\mu^a \delta A_\nu^b}
      \right\rangle \nonumber \\
  &=  
       \frac{-i}{2V}  \ln \det
         \left[  
             g^{\mu\nu} \delta^{ab} 
             \bigl( -p^2 + 9 g^2  \phi_{\rm g} \bigr)
         \right] \nonumber \\
  &=  -16i \int \frac{\md^4 p}{(2\pi)^4}
       \ln \left[  
             -p^2 + \Mg^2
             \right],
\end{align}
with the effective gluon mass $\Mg^2 = 9g^2\pg$.
Then the effective potential can be combined as
\begin{align}
  {\Omega}_{\rm g}
  = 72 g^2 \phi_{\rm g}^2
      -16i \int \frac{\md^4 p}{(2\pi)^4}
       \ln \left[  -p^2 + \Mg^2 \right],
\end{align}
and its finite temperature ($T$) and chemical potential ($\mu$)
extension is given by
\begin{align}
  {\Omega}_{\rm g}
  &= 72 g^2 \phi_{\rm g}^2 \nonumber \\
  &+16 \int \frac{\md^3 p}{(2\pi)^3}
       \left[  \eg 
               +T \sum_{\pm} \ln(1-e^{-\beta \eg^\pm}) 
       \right],
\label{eq:g_pot_t}
\end{align}
where $\eg^\pm = \eg \pm \mg$ with the chemical potential
for gluons and $\eg = \sqrt{p^2 + \Mg^2}$. The brief derivation
of the effective potential is presented in Appendix~\ref{app:pot}.

In this model, we treat the gluon condensate as the order parameter
whose expectation value is determined by searching the minimum of
the effective potential, 
\begin{align}
  \frac{\partial {\Omega}_{\rm g}}{\partial \pg} = 0.
\end{align}
The solution corresponds to the stational point of the effective
potential, and the obtained equation can be reduced to
\begin{align}
  \phi_{\rm g}
  = \int \frac{\md^4 p}{(2\pi)^4}
     \frac{-i}{p^2-\Mg^2}.
\label{eq:gap_g0}
\end{align}
This indicates the one-loop amplitude of the effective
gluon propagator,
\begin{align}
  D^{\rm g}_{\rm eff} (p)
  = \frac{-i}{p^2-\Mg^2}.
\label{eq:prop_g}
\end{align}
with the effective gluon mass $\Mg$. Thus the equation forms
the self consistent system. The finite temperature
and chemical potential extension of Eq.~(\ref{eq:gap_g0})
becomes
\begin{align}
  \phi_{\rm g}
  = -\int \frac{\md^3 p}{(2\pi)^3}
     \frac{1}{2\eg} 
     \left[ 1 - \sum_\pm \frac{1}{e^{\beta \eg^\pm} -1}
     \right].
\label{eq:gap_g}
\end{align}
This is the gap equation  for the gluon condensate
which we will numerically solve in the next section.

\subsection{\label{subsec:q_gap}%
Quark gap equation}
To obtain effective quark contribution, we fist consider
the four-fermion interaction in the two-flavor system
following the discussion given in~\cite{Kohyama:2016dhd}.

The QCD Lagrangian density is written by
\begin{align}
 & \mathcal{L}_{\rm QCD}
   =  \mathcal{L}_{\rm q}^0 
    + \mathcal{L}_{\rm I}
    + \mathcal{L}_{\rm g}, \\
 & \mathcal{L}_{\rm q}^0
   = \overline{q} (i\partial\!\!\!/ -m) q,\\
 & \mathcal{L}_{\rm I}
      = g \overline{q} \gamma^\mu t^a q A^a_\mu,
\end{align}
The partition function can be expanded by using the Taylor
series as
\begin{align}
  & \mathcal{Z}_{\rm QCD}
   =
    \int \! \mathcal{D}q \int \! \mathcal{D}\! A 
    \exp \left[
      i\int \md^4 x 
      \mathcal{L}_{\rm QCD}
       \right] \nonumber \\
   & =
      \int \! \mathcal{D}q \int \! \mathcal{D}\! A 
      e^{i\int \md^4 x{\mathcal L}_0} 
      \sum_{n=0}^{\infty}
      \frac{1}{n!}
      \left(  i \int \md ^4 x  \mathcal{L}_{\rm I} 
      \right)^n,
\end{align}
with ${\mathcal L}_0 = {\mathcal L}_{\rm q}^0 + {\mathcal L}_{\rm g}$.
Here we are interested only in the quark contribution, then focus on the
amplitude,
\begin{align}
   &\mathcal{Z}_{\rm q}
     \propto
      \int \! \mathcal{D}q \int \! \mathcal{D}\! A 
      e^{i\int \md^4 x{\mathcal L}_{\rm q}^0} 
      \biggl[  1 
         +  \frac{1}{2} 
              \left( ig \int \md ^4 x  
                       {\mathcal L}_{\rm I}  \right)^2 
      \biggr].
\label{eq:Z_QCD}
\end{align}
The explicit form of the above equation is written by
\begin{align}
   &\mathcal{Z}_{\rm q}
     \propto
      \int \! \mathcal{D}q \int \! \mathcal{D}\! A 
      e^{i\int \md^4 x{\mathcal L}_{\rm q}^0} \nonumber \\
   & \quad \times
      \biggl[  1 + 
                 \frac{1}{2} \left( ig \int \md ^4 x  
                 \overline{q} \gamma^\mu t^a q A^a_\mu  
      \right)^2
      \biggr].
\label{eq:Z_g4}
\end{align}
The four-fermion contact interaction can be derived by assuming
the gluons have sharp momentum peak around some specific
energy scale in the propagator,
\begin{align}
      \bigl\langle
           A^a_{\mu} (x)  
           A^b_{\nu} (y)
      \bigr\rangle_{\rm p}
      =
      \int \frac{\md^4 p}{(2\pi)^4}
      \frac{-ig_{\mu \nu} \delta^{ab}}{p^2 + i\epsilon}
      e^{-i p \cdot (x-y)}.
\end{align}
The replacement $p^2 \to \etag^2$ leads
\begin{align}
      \bigl\langle
           A^a_\mu (x)  
           A^b_\nu (y)
      \bigr\rangle_{\rm const}
      =
      \frac{-ig_{\mu \nu} \delta^{ab} }{\etag^2}      
      \delta^{(4)} (x-y),
\label{eq:replace}
\end{align}
and it induces the four-fermion contact interaction 
\begin{align}
   &\mathcal{Z}_{\rm q}
     \simeq
      {\mathcal N}_A \int \! \mathcal{D}q 
      e^{i\int \md^4 x{\mathcal L}_0} \nonumber \\
   & \quad \times
      \biggl[  1 + 
        \frac{i g^2}{2 \eta_{\rm g}^2} 
        \int \md ^4 x  
        \left( \overline{q} \gamma^\mu t^a q  \right)  
        \left( \overline{q} \gamma_\mu t^a q  \right)  
      \biggr].
\label{eq:Z_q}
\end{align}
with the over all factor ${\mathcal N}_A$ with respect to the gluon
integration. Putting back the resulting term into the exponential
using $1+\epsilon \simeq e^{\epsilon}$, we have the following
effective Lagrangian for quarks
\begin{align}
   &\mathcal{L}_{\rm eff}
     =
     \overline{q} (i\partial\!\!\!/ -m) q
     + 
        \frac{g^2}{2 \etag^2} 
        \left( \overline{q} \gamma^\mu t^a q  \right)  
        \left( \overline{q} \gamma_\mu t^a q  \right).
\label{eq:L_eff}
\end{align}
This is how we derive the four-fermion contact interaction based
on QCD with condensed gluons.

The applications of  the mean field approximation,
$\phi_{\rm q} \simeq \langle \bar{q}q \rangle$, after the Fiertz
transformation gives
\begin{align}
 & \tilde{\mathcal{L}}_{\rm q}
   = \bar{q} (i\partial\!\!\!/ -\hat{M}) q
     -  2G(\phi_{\rm u}^2 +\phi_{\rm d}^2),
\end{align}
where $G = 2g^2/(9\etag^2)$ and
$\hat{M}$ is the diagonal
matrix with $M_{\rm q} = m_{\rm q} -4G \phi_{\rm q}$. One can
evaluate the effective potential from the above mean-field form,
\begin{align}
  \Omega_{\rm q}
  &= 2G(\phi_{\rm u}^2 +\phi_{\rm d}^2) \nonumber \\
     &-2N_c \sum_{\rm q}
           \int \frac{{\rm d}^3q}{(2\pi)^3}
           \left[ E_{\rm q}
           + T \sum_{\pm}     
           \ln \Bigl(1 + e^{-\beta E_{\rm q}^{\pm}} \Bigr)
           \right],
\label{eq:q_pot_t}
\end{align}
with the number of colors $N_c(=3)$,
$E_{\rm q}^\pm = E_{\rm q} \pm \mq$
and $E_{\rm q} = \sqrt{q^2 + M_{\rm q}^2}$. From the stable
condition of the effective potential, one has the gap equation,
\begin{align}
 \phi_{\rm q}=
     -{\rm tr} \int \! \frac{\md^4 q}{(2\pi)^4}
          \frac{i}{q\!\!\!/ - M_{\rm q}},
\end{align}
where the right hand side expresses the one-loop amplitude of the
propagator for quarks
\begin{align}
 S^{\rm q}_{\rm eff} (q)=
          \frac{i}{q\!\!\!/ - M_{\rm q}},
\end{align}
as seen in the gluon case.
The finite temperature and chemical potential form becomes
\begin{align}
 \phi_{\rm q}=
     -4N_c M_{\rm q} \!\! \int \! \frac{\md^3 q}{(2\pi)^3}
          \frac{1}{2E_{\rm q}}
     \left[ 1 -
       \sum_\pm \frac{1}{e^{\beta E_{\rm q}^{\pm}}+1}
   \right].
\label{eq:q_gap}
\end{align}
As frequently studied in quark effective model analyses, this gives
the expectation values of the chiral condensates. Although this model
is the same with the NJL model, we call it as the pure quark model
in this letter to explicitly distinguish the model with gluon condensate.

\subsection{\label{subsec:model}%
Coupled equations}
We have evaluated the effective potential both for the gluon and
quark condensates above. One should note that the obtained
forms compose the decoupled system since we set the effective
coupling strength for four-fermion interaction as the constant.
However, it is expected that the effective four-fermion interaction
becomes weak at high temperature where the gluon condensate is
as well expected to be small. Then we modify the form for the
four-point coupling so that one can construct more reliable
effective theory.

The constant property comes from the Eq.~(\ref{eq:replace})
in which the expectation value $\langle A^a_\mu A^b_\nu \rangle$
set to be constant. Here we modify this relation by
\begin{align}
      \bigl\langle
           A^a_\mu (x)  
           A^b_\nu (y)
      \bigr\rangle
      =
     -i g_{\mu \nu} \delta^{ab}  \cdot c \pg
      \delta^{(4)} (x-y),
\label{eq:replace_co}
\end{align}
with the constant $c$, which enables us to incorporate the
temperature dependence into the effective coupling.
Note that the treatment in Eqs.~(\ref{eq:g_cond}) and (\ref{eq:replace})
is essentially different; the former is obtained after the renormalization,
and the latter still keeps the divergent contribution at $y=x$. The relation
between these quantities are non-trivial, therefore we introduce the
coefficient $c$ in Eq.~(\ref{eq:replace_co}).
Determining the coupling structure, we see the effective potential of
the whole system,
\begin{align}
  {\Omega}
    &= 72 g^2 \phi_{\rm g}^2
      + \frac{4}{9} g^2 c \pg (\phi_{\rm u}^2+\phi_{\rm d}^2)
      \nonumber \\
    &+16 \int \frac{\md^3 p}{(2\pi)^3}
       \left[  \eg 
               +T \sum_{\pm} \ln(1-e^{-\beta \eg^\pm}) 
       \right]
     \nonumber \\
    &-6 \sum_{\rm q}
           \int \frac{{\rm d}^3q}{(2\pi)^3}
           \left[ E_{\rm q}
           + T \sum_{\pm}     
           \ln \Bigl(1 + e^{-\beta E_{\rm q}^{\pm}} \Bigr)
           \right],
\label{eq:pot_c}
\end{align}
where $M_{\rm q} = m_{\rm q} -8/9 \,  g^2 c \pg \phi_{\rm q}$
appearing in the quasi-particle energy for quarks, $E_{\rm q}$.
The coupled gap equations are calculated by the following
simultaneous condition,
\begin{align}
  \frac{\partial {\Omega}}{\partial \pg} = 0,
  \quad
  \frac{\partial {\Omega}}{\partial \phi_{\rm q}} = 0.
\end{align}
The above conditions give the explicit forms,
\begin{align}
  \phi_{\rm g}
  = &-\frac{c}{324}(\phi_{\rm u}^2+\phi_{\rm d}^2)
       -\int \frac{\md^3 p}{(2\pi)^3} \frac{1}{2\eg} 
        \left[ 1 - \sum_\pm 
                 n(\eg^\pm)
        \right] \nonumber \\
      &-\frac{c}{27} \sum_{\rm q}
        \int \frac{\md^3 q}{(2\pi)^3}
        \frac{\phi_{\rm q}}{2E_{\rm q}} 
        \left[ 1 - \sum_\pm 
                 f(E_{\rm q}^\pm)
        \right],
\label{eq:gap_cg} \\
  \phi_{\rm q}
  = &-4N_c M_{\rm q} \!\! \int \!
        \frac{\md^3 q}{(2\pi)^3} \frac{1}{2E_{\rm q}}
        \left[ 1 -\sum_\pm f(E_{\rm q}^\pm) \right],
\label{eq:gap_cq}
\end{align}
with the distribution functions, $n(E) = ({e^{\beta E} -1})^{-1}$
and $f(E) = ({e^{\beta E} +1})^{-1}$, for bosons and fermions.
These equations will give the expectation values of the gluon
and chiral condensates.

\section{\label{sec:separate}%
Separate models}
We have formulated the effective models for the gluon
and chiral condensates in the previous section. We will now
present actual numerical analyses concerning on the pure
gluon and quark models.

\subsection{\label{subsec:nc_g}%
Pure gluon model}
Since the one-loop integral in the gap equation diverges, we
need to introduce some regularization procedure to obtain finite
model predictions.
It is important to note that the gluon condensate becomes negative
if we apply the three- or four-momentum cutoff schemes as usually
introduced for the pure quark model, which leads the negative mass
square $\Mg^2 < 0$. The negative mass square behaves badly in
calculating the thermodynamical quantities, so we shall employ
the dimensional regularization prescription which gives the positive
gluon condensate~\cite{Inagaki:2015lma}, namely the positive
mass square. The more detailed discussions on the sign of the
condensate and the regularization prescriptions are presented
in Apps.~\ref{app:sign} and \ref{app:reg}. In the model with the
dimensional regularization the gluon part has three parameters,
the spacetime dimension $D$ in the loop integral, the mass scale
$M_0$, and the coupling constant for the strong interaction $g$.
The coupling is chosen to be $g=1.12$ from the consideration on
the quark sector with $\etag=225$MeV, which will be discussed later.
As for the remaining parameters, we set $D=2.17$ and $M_0=67$MeV
so that we have $M_g=225$MeV($=\etag$) at zero temperature and
the critical temperature $T^{\rm g}_c=270$MeV for pure gluon sector.

\begin{figure}[h!]
\begin{center}
   \includegraphics[width=7.5cm,keepaspectratio]{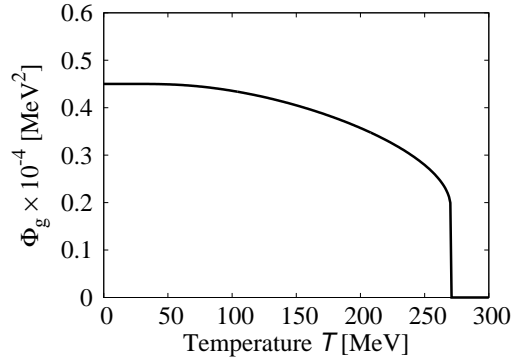}
   \caption{\label{fig:PM_g}
   The gluon condensate.
   }
\end{center}
\end{figure}

Figure \ref{fig:PM_g} displays the numerical results of the gluon
condensate.
One sees that the condensate gradually decreases for low temperature,
then suddenly drops around $T=270$MeV. It should be mentioned that
the gap equation allows a negative value for the solution,
in which we regard unphysical mode so we set $\pg=0$ for the
region corresponding to the case with the negative condensate.
From the plots of the gluon condensate, we see that the curve actually
shows the tendency of the symmetry restoration at finite temperature,
which, we think, is the desired solution of the gap equation for the gluon condensate, Eq.~(\ref{eq:gap_g}).

\subsection{\label{subsec:nc_q}%
Pure quark model}
When one sets the effective coupling strength for four-fermion interaction
to be a constant, the model reduces to the NJL model as mentioned above,
and which has five parameters: the gluon energy scale $\etag$,
three momentum cutoff $\Lambda_{\rm q}$, the coupling constant for the
strong interaction $g$, the current quark masses $m_{\rm u}$ and
$m_{\rm d}$. We
consider the isospin symmetric case $m_{\rm u}=m_{\rm d}(=m_{\rm q})$
since the mass difference between the up and down quarks is
small comparing to the hadronic scale. Here we
chose the value $m_{\rm q}=5.5$MeV for the current quark mass,
and $\etag=225$MeV for the gluon energy scale which is from
the proton radius $r_{\rm P} \simeq 0.87{\rm fm}$. We then fix
the remaining parameters by $\Lambda_{\rm q}=631$MeV and
$g=1.12$ so that the model reproduces
the observed pion mass and decay constant, $m_{\pi}=138$MeV and
$f_{\pi}=93$MeV following \cite{Hatsuda:1994pi}.

\begin{figure}[h!]
\begin{center}
   \includegraphics[width=7.5cm,keepaspectratio]{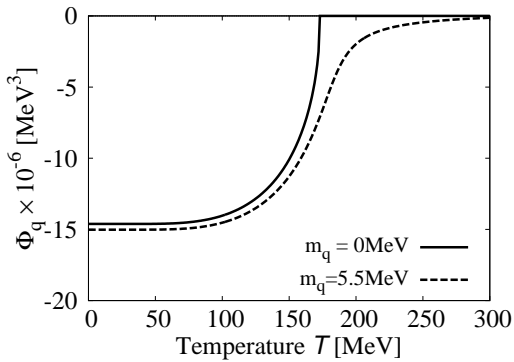}
   \caption{\label{fig:PM_q}
   The chiral condensate for $m_{\rm q}=0$ and $5.5$MeV
   with other parameters fixed.
   }
\end{center}
\end{figure}

The numerical results of the chiral condensates for
current quark mass $m_{\rm q}=0$ and $5.5$MeV are
shown in Fig.~\ref{fig:PM_q}.
These are frequently drawn curves calculated in a lot of preceding
works, where the absolute value of the chiral condensate decreases
at high temperature. The curve becomes smoother for the massive
current quark mass case, which comes from the effect of the explicit
symmetry breaking. The critical temperature of the chiral phase
transition (defined by the maximum change of the condensate with
respect to $T$ and/or $\mu$) for the above parameters set
becomes $T_c^{\rm q}=178$MeV, being close to the expected value
from the lattice QCD simulations,
$T^{\rm q}_c \simeq 175$MeV~\cite{Aoki:2006br}.

\subsection{\label{subsec:comparison}%
Comparisons with BEC and BCS}
It may also be interesting to compare the results of the current
model and the actual experimental data of the BEC state and the
prediction from the BCS theory.

\begin{figure}[h!]
\begin{center}
   \includegraphics[width=7.5cm,keepaspectratio]{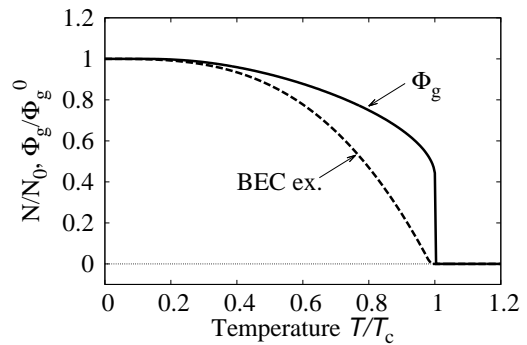}
   \caption{\label{fig:ratio_g}
   Comparison with the fit of the BEC experimental data
   for the condensate atoms. The dashed curve is drawn by
   using the fitted formula, Eq (\ref{eq:bec_fit})~\cite{Henn:2008}.
   }
\end{center}
\end{figure}

Figure \ref{fig:ratio_g} plots the temperature dependence on the
gluon condensate and the condensed atoms in the BEC state.
The dashed curve is the fitted result
\begin{equation}
   \frac{N_0}{N} = 1-\left(\frac{T}{T_c}\right)^3
   -4a \left( \frac{T}{T_c} \right)^{7/2},
\label{eq:bec_fit}
\end{equation}
with $a=0.01$,  where the transition temperature is
$T_c = 142$nK for 87Rb~\cite{Henn:2008}. One confirms that
the two results basically show the resemblance, while curves a
bit deviate. The deviation arises due to the difference between
fitted formula and our gap equation. We note that the condensate
drops rapidly at high temperature around $T_c$, which comes
from the characteristic property of the Bose-Einstein distribution
function, $n(E)$. The term can become infinitely large for
$e^{\beta \eg} \simeq 1$, then causes crucial effect on the
solution at high temperature. This is the numerical reason on the
drastic behavior of the condensate and its deviation from the BEC
data at high temperature.

Although there exists the deviation from the expected
behavior, we still think that the gap equation captures the
phenomena of the phase transition at the satisfactory level for
practical usage.
We then regard Eq.~(\ref{eq:gap_g}) as the effective gap equation
for the gluon condensate, then proceed the calculations on various
thermodynamic quantities based on it.

\begin{figure}[h!]
\begin{center}
   \includegraphics[width=7.5cm,keepaspectratio]{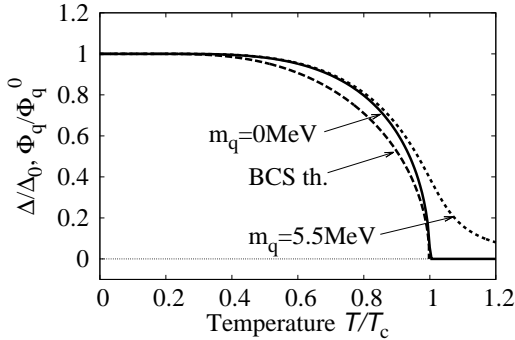}
   \caption{\label{fig:ratio_q}
   Comparison with the result from the BCS theory for the gap
   equation, Eq~(\ref{eq:bcs_gap}), in superconductivity (dashed curve).
   The solid and dotted curves exhibit the results in the pure quark
   model for $m_u=0$ and $5.5$MeV with the other parameters fixed.
   }
\end{center}
\end{figure}

We can as well check the similarity between the results of
the chiral condensate and the BCS theory for superconductivity
as seen in Fig.~\ref{fig:ratio_q}.
The solid and dotted plots are the obtained chiral condensate
in the pure quark sector for $m_{\rm q}=0$ and $5.5$MeV,  and
the dashed one is the prediction from the BCS gap equation,
\begin{align}
 \frac{\hbar \omega}{Nv}
 =  \int_{0}^{\hbar \omega} \!\! \md \xi \,
    \frac{1}{E_\xi} 
    \left[ 1-\frac{2}{e^{\beta E_\xi}+1} \right],
\label{eq:bcs_gap}
\end{align}
where $N$ is the number density, $v$ is the potential of the system,
and $E_{\xi}=\sqrt{\xi^2 + \Delta^2}$ with the BCS gap energy
$\Delta$~\cite{Tinkham:1975}. The parameters in the BCS gap
equation are chosen as $Nv =0.3 \hbar \omega$ with
$\hbar \omega = 8.98 \times 10^{-3}$eV
so that we have $T_c=4.2$K for Hg. One notes that the lines of the
chiral condensate for massless case and the BCS gap energy are
very close as seen in the figure. This is due to the similarity
of the quark gap equation and the BCS gap equation as obviously
confirmed by Eqs.~(\ref{eq:q_gap}) and (\ref{eq:bcs_gap}).
The model with massive current quark gives
smoother decrease with respect to $T$ since the chiral
condensate can not be restored
completely with the non-zero current quark mass.

We have thus seen that both the gluon and chiral condensates
resemble with the behavior of the BEC and BCS states.  We believe
this is indeed natural since our original motivations for the model
constructions come from the analogies with these two typical
phenomena in condensed matter physics.

\section{\label{sec:coupled}%
Coupled model}
We have tested the condensates through the two gap equations
separately in the previous section. Let us now discuss what happens
when we consider the coupled system based on the simultaneous
equations, Eqs.~(\ref{eq:gap_cg}) and (\ref{eq:gap_cq}).

The coupled model has six parameters:
the current quark masse $m_{\rm q}$,
three-momentum cutoff $\Lambda_{\rm q}$,
coupling constant for the strong interaction $g$,
spacetime dimensions $D$, the mass scale $M_0$
and effective coupling coefficient $c$. We set these
values by
\begin{align*}
  &m_{\rm q}=5.5{\rm MeV}, \quad
    \Lambda_{\rm q} = 631{\rm MeV}, \quad
    g = 1.12, \\
  &D = 2.17, \quad
    M_0 =67{\rm MeV}, \quad
    c = 0.439 \times 10^{-8} {\rm MeV}^{-4}.
\end{align*}
Here we again present the reasoning on the above parameter
choices. $m_{\rm q}$ is guessed from the data by
Particle Data Group~\cite{Agashe:2014kda} around the
scale $\sim 1$GeV. $\Lambda_{\rm q}$ and $g$
are fitted so that the model leads the observed pion mass and
decay constant $m_\pi=138$MeV and $f_{\pi}=93$MeV
in the pure quark model following~\cite{Hatsuda:1994pi}.
$D$ and $M_0$ are set so as to produce the values $\Mg=225$MeV
at zero temperature and $T^{\rm g}_c=270$MeV in the pure gluon
model. $c$ is chosen to make the effective quark mass to be
$M_{\rm q}=335$MeV at $T=0$ and $\mu=0$ in the coupled model.
With these fitted parameters, we shall be studying the
coupled system through solving the gap equations.

\subsection{\label{subsec:cond}%
Gluon and chiral condensates}
\begin{figure}[h!]
\begin{center}
   \includegraphics[width=7.5cm,keepaspectratio]{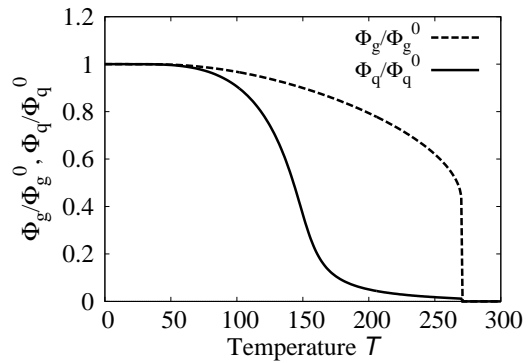}
   \caption{\label{fig:ratio_c}
   Gluon and chiral condensates normalized by
   $\pg^0=(67{\rm MeV})^2$ and $\pq^0=(-247{\rm MeV})^3$.
   }
\end{center}
\end{figure}

Figure \ref{fig:ratio_c} shows the obtained solutions for the
gluon and chiral condensates.
We note that the decrease of the chiral condensate starts
at lower $T$ compared to the pure quark model, while the solution
for the gluon condensate is not affected in the coupled system.
This can be understood by the following observations; the
effective coupling for the four-fermion interaction gets weak
at high temperature since the gluon condensate is smaller for
high $T$, which lowers the critical temperature of the chiral
phase transition. While the
corresponding coupling on the gluon part is constant, and does
not relate to the strength of the chiral condensate.  Therefore, the 
critical behavior of the gluon condensate can not be affected by
the quark part.

The obtained solutions may be reasonable in the current model
study where the underlying assumption is that the quark
condensate coming from the effective four-fermion interaction
is driven by the gluon condensate. Considering the above
mentioned property of the coupled equations, we will pay the
attention to the difference on the quark system between the
coupled and pure quark models in what follows.

\subsection{\label{subsec:comp}%
Comparison with pure quark model
}
We see that the chiral condensate is influenced by the gluon
condensate through the four-point interaction. It may be
interesting to compare how the chiral condensate changes
due to the gluon fields in the coupled system.

\begin{figure}[h!]
\begin{center}
   \includegraphics[width=7.5cm,keepaspectratio]{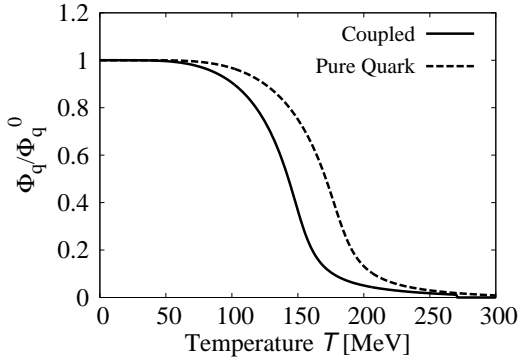}
   \caption{\label{fig:ratio_comp}
   The chiral condensates.
   }
\end{center}
\end{figure}

The influence of the gluon condensate effect is shown
in Fig. \ref{fig:ratio_comp},
where the decrease starts at the lower temperature in the coupled
model. As mentioned above, this is straightforward consequence
of the weakened effective coupling. The critical temperature for
the coupled model and pure quark model are $T_c^{\rm q}=149$MeV
and $178$MeV, respectively. Thus we numerically confirm the
change of the critical temperature being around $30$MeV.

\subsection{\label{subsec:pressure}%
Pressure}
It may be intriguing to study the pressures which are defined by
\begin{align}
  &\mathcal{P}_{\rm g} =
  -\bigl[ \Omega_{\rm g}(T,\mu) - \Omega_{\rm g}(0,0) \bigr], \\
  &\mathcal{P}_{\rm q} =
  -\bigl[ \Omega_{\rm q}(T,\mu) - \Omega_{\rm q}(0,0) \bigr].
\end{align}
So they are set by the difference of the effective potential from the
value at $T=0$ and $\mu=0$.

\begin{figure}[h!]
\begin{center}
   \includegraphics[width=7.5cm,keepaspectratio]{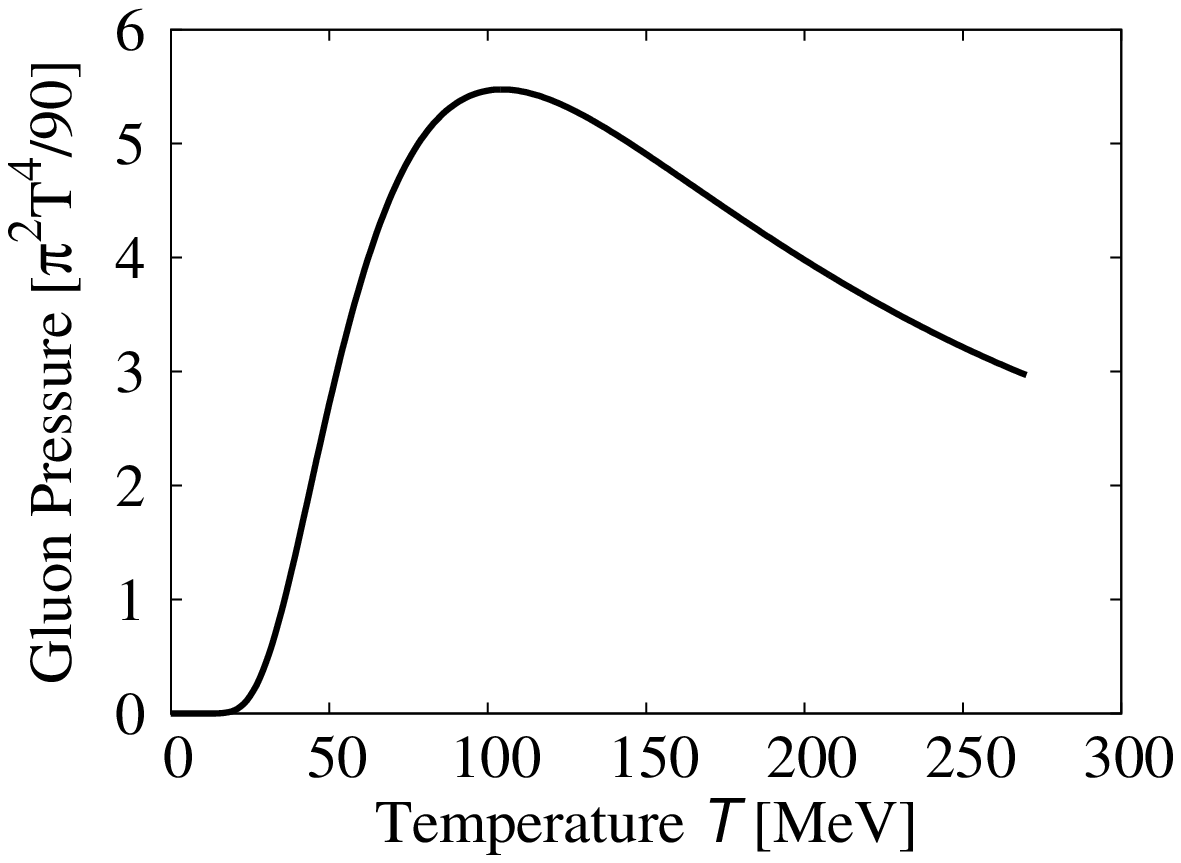}
   \includegraphics[width=7.5cm,keepaspectratio]{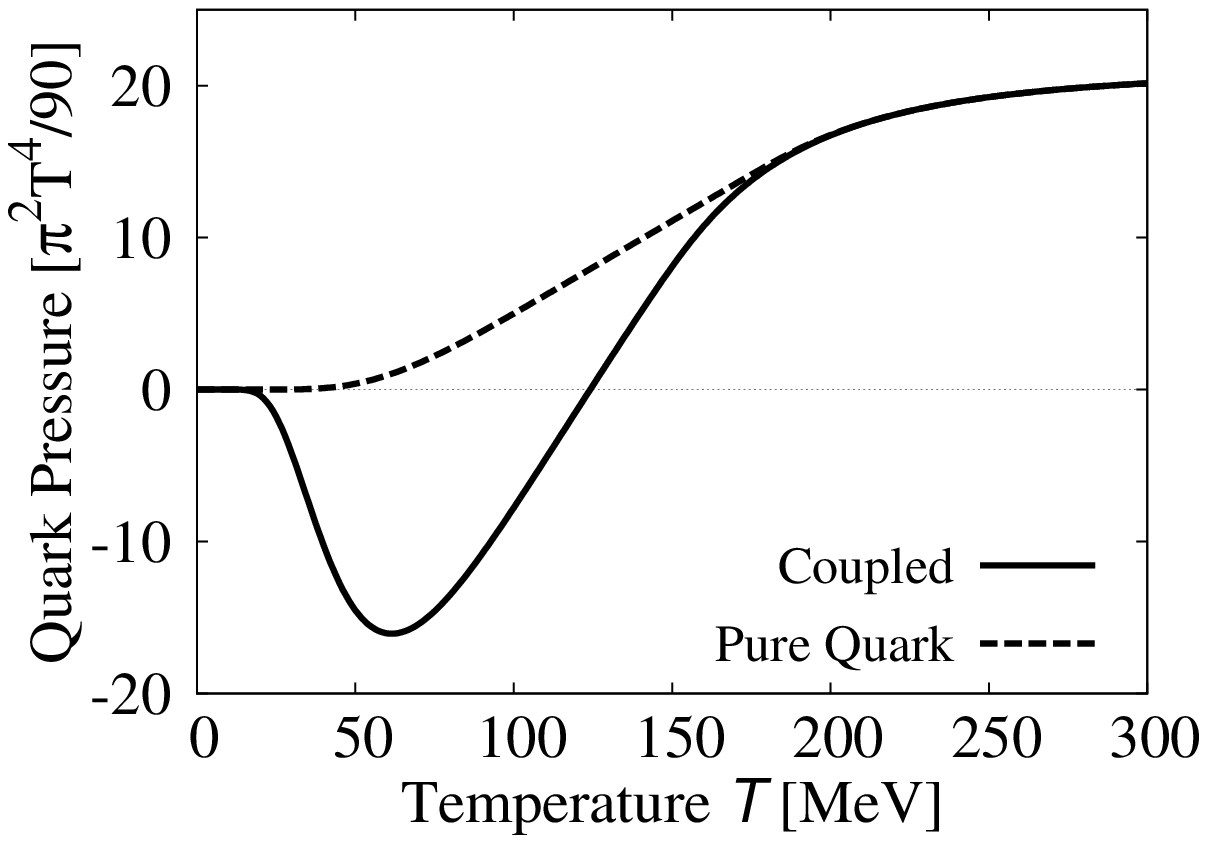}
   \caption{\label{fig:pressure}
   Gluon and quark pressures.
   }
\end{center}
\end{figure}

The pressure from the gluon sector in the coupled models is
displayed in the upper panel of  Fig. \ref{fig:pressure}.
It should be noted that the gluon sector generates the same curve
in the coupled and pure gluon models as discussed in
Sec. \ref{subsec:cond}, therefore we only show the result of the
coupled model for the gluon part. It is as well to be noted that the
gap equation leads the unphysical solution with the negative mass
square above $T_c^{\rm g}=270$MeV, there we do not show the
result since the thermodynamic quantities are not well defined
above $T_c$ in the current model analyses. The pressure of the gluon
contribution normalized by $\pi^2T^4/90$ increases for low $T$,
then decreases at high temperature.
The decrease at high $T$ numerically comes from our choice of
the dimensional regularization method where the dimensions in
the integral is smaller than four, then the contribution from the
temperature dependent term is suppressed.

The interesting difference is seen in the lower panel of 
Fig. \ref{fig:pressure} where the pressures from the quark sector
are plotted both in the coupled and pure quark models.
The contribution becomes negative for low $T$ and rises up for
high $T$ in the coupled model, while it monotonically increases
with temperature in the pure quark model. This may indicate that
the background gluon condensate tends to suppress the quark
excitations in the confined state. On the other hand above
$T_c^{\rm q}$, these two models produce the same results since
there the effective quarks turns into current quarks and the
temperature contribution dominates the system.

\subsection{\label{subsec:susceptibility}%
Susceptibilities}
The number density and susceptibility evaluated by
\begin{align}
  &N_{\rm g} = -\frac{\partial \Omega_{\rm g}}{\partial \mg}, \quad
  N_{\rm q} = -\frac{\partial \Omega_{\rm q}}{\partial \mu_{\rm q}}, \\
  &\chi_{\rm g} = -\frac{\partial^2 \Omega_{\rm g}}{\partial \mgs}, \quad
  \chi_{\rm q} = -\frac{\partial^2 \Omega_{\rm q}}{\partial \mu_{\rm q}^2},
\end{align}
are important thermodynamic quantities which we will present in
this subsection.

\begin{figure}[h!]
\begin{center}
   \includegraphics[width=7.5cm,keepaspectratio]{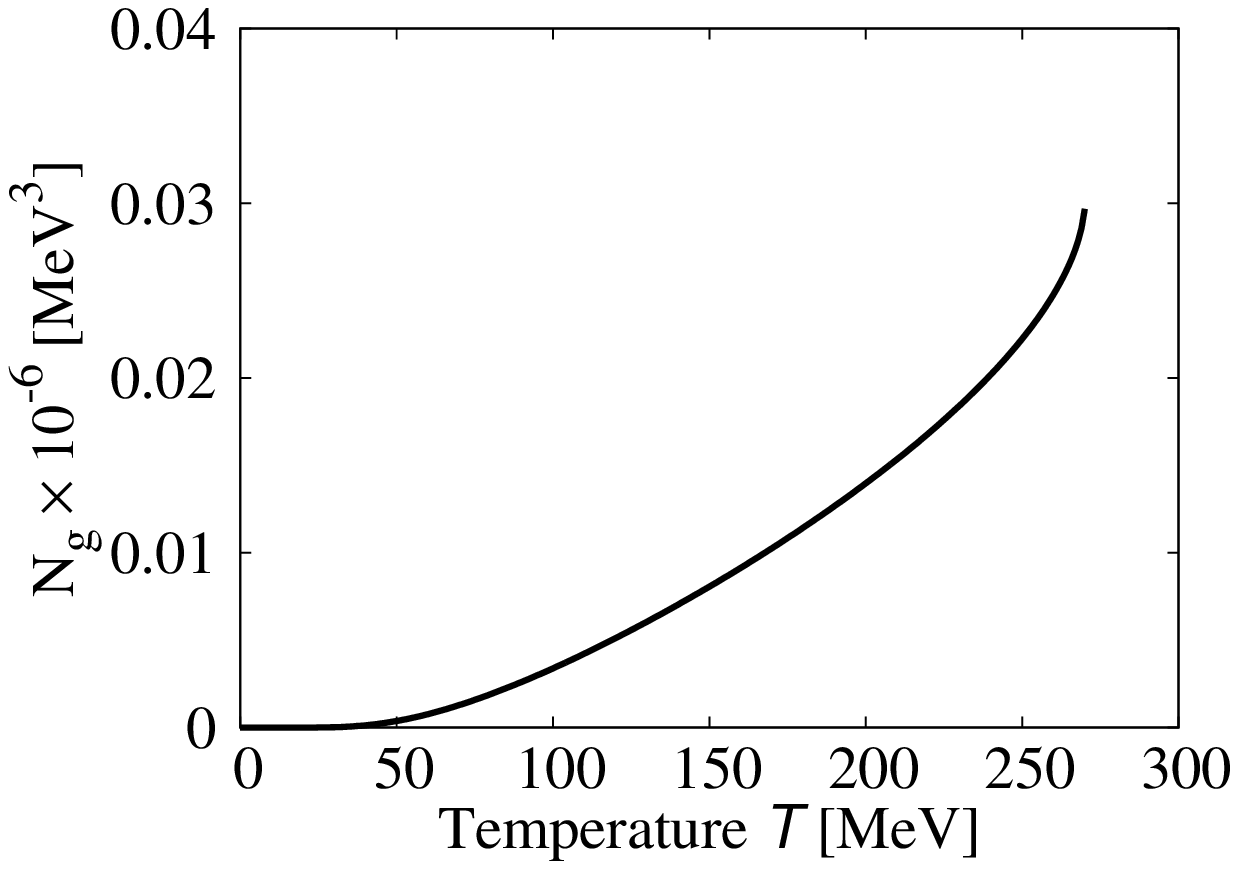}
   \includegraphics[width=7.5cm,keepaspectratio]{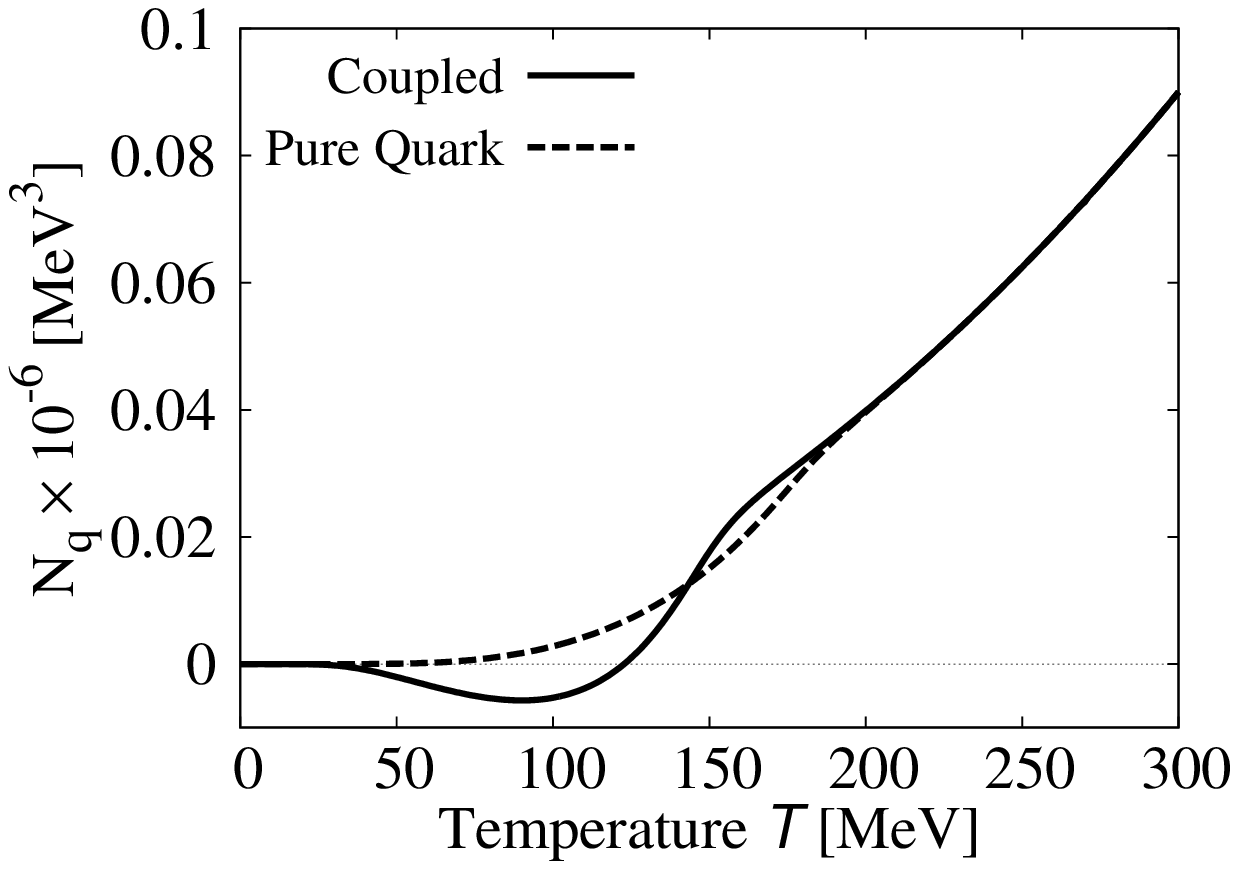}
   \caption{\label{fig:number}
   Gluon and quark number densities.
   }
\end{center}
\end{figure}

Figure \ref{fig:number} plots the number density  for the gluon and
quark sectors. 
It is seen that the curves for gluon sector and pure quark model
indicate monotonic increase, while $N_{\rm q}$ in the coupled
model exhibits different feature; the density becomes negative for
low $T$ then gets larger for high $T$. Although the negative number
density seems to be unphysical which comes from the negative
pressure, we get positive density being close to the one from the
pure quark model above $T_c^{\rm q} \simeq 150$MeV as has
been expected.

\begin{figure}[h!]
\begin{center}
   \includegraphics[width=7.5cm,keepaspectratio]{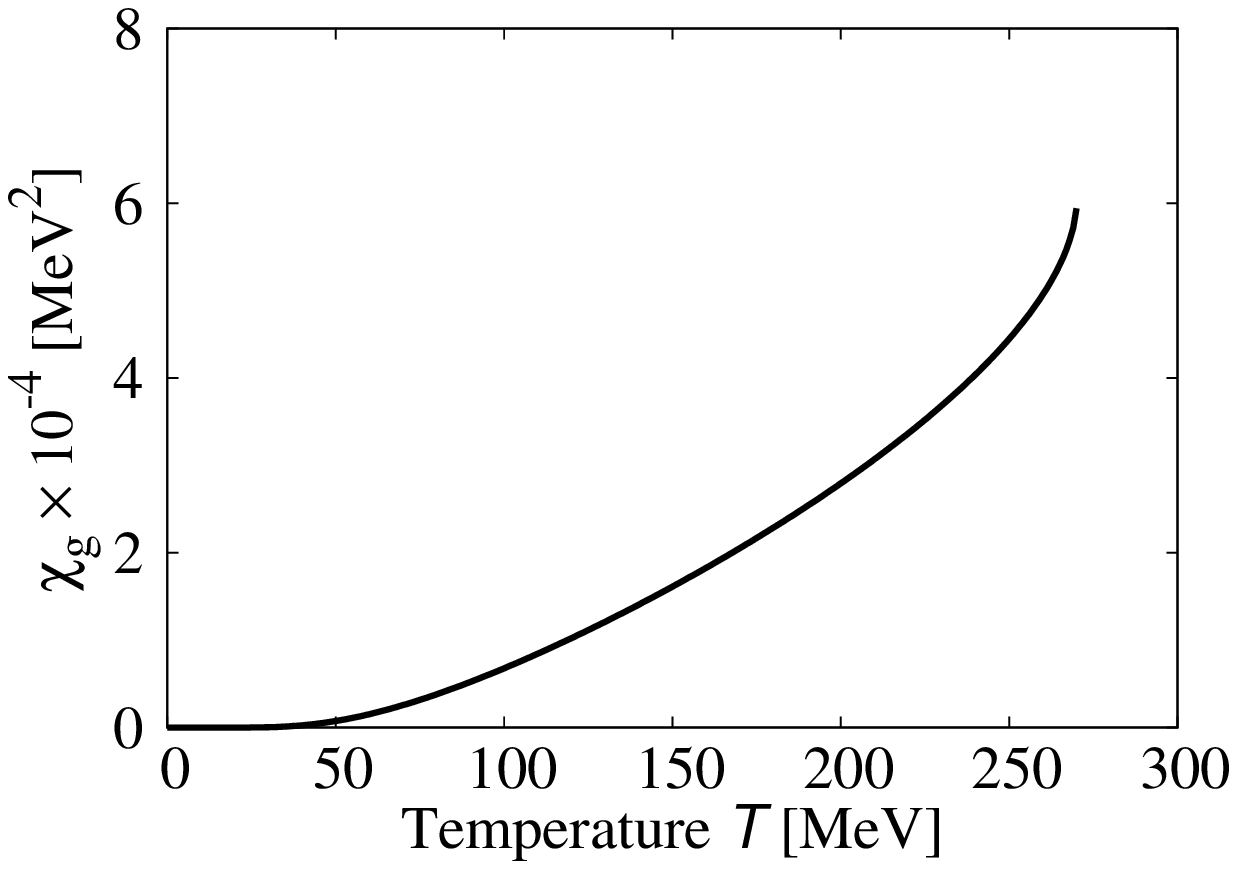}
   \includegraphics[width=7.5cm,keepaspectratio]{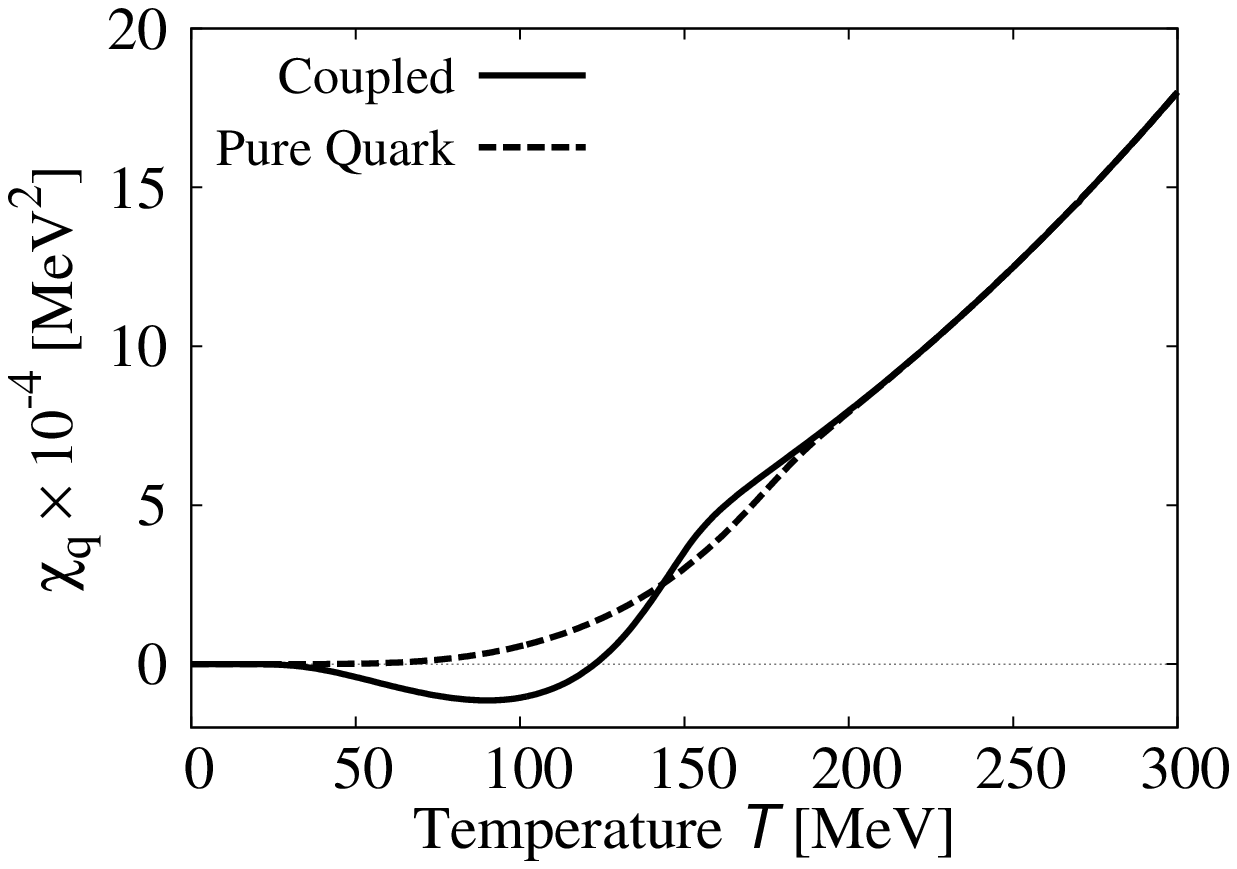}
   \caption{\label{fig:sus}
   Gluon and quark number susceptibilities.
   }
\end{center}
\end{figure}

The susceptibilities for the gluon and quark sectors are
displayed in the upper and lower panels in Fig. \ref{fig:sus}.
One notes a similar tendency on these results; the gluon
sector and pure quark model draw the monotonically increasing
lines, and the coupled model has negative value for low $T$.

We thus find the non-monotonic structure on the number
density and susceptibility for the quark sector in the coupled
model; the quark excitations are suppressed at low temperature.
This is basically due to the change of the effective four-fermion
coupling in the coupling system.

\section{\label{sec:chiral_phase}%
Chiral phase transition}
We have seen the effect of the gluon condensate on the chiral
condensate at finite temperature in the previous section. We think
now it is interesting to study the system with finite quark chemical
potential, in particular, the phase structure on the chiral phase
transition.

\subsection{\label{subsec:pd}%
Phase diagram of chiral condensate}
\begin{figure}[h!]
\begin{center}
   \includegraphics[width=7.5cm,keepaspectratio]{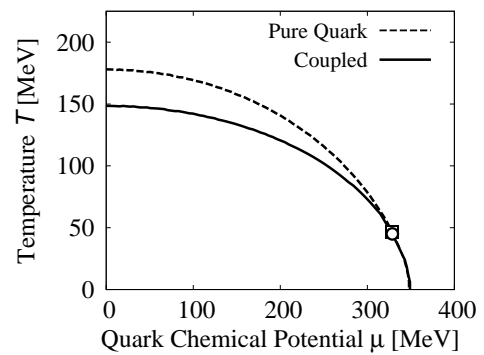}
   \caption{\label{fig:pd}
   Phase diagram for the chiral condensate. The circle (square)
   represents the critical point in the coupled (pure quark) model.
   }
\end{center}
\end{figure}

We show how the phase transition of the chiral condensate has
influence from the back ground gluon condensate in Fig.~\ref{fig:pd}.
The region of the broken phase shrinks toward the lower temperature
direction when one considers the coupled model, and there does not
appear the noticeable difference on the critical chemical potential for
low $T$. This feature can easily be understood, since the gluon condensate
becomes smaller at finite temperature. On the other hand, it is expected
not to receive drastic change from the finite quark chemical potential
$\mu_{\rm q}$, then the transition chemical potentials are the same
between these models. It may be worth aligning the actual values of
the critical chemical potential and temperature for two models. The
coupled model gives $(\mu_{\rm CP}^{\rm q}, T_{\rm CP}^{\rm q}) 
= (329{\rm MeV},45{\rm MeV})$, and the pure quark model does
$(\mu_{\rm CP}^{\rm q}, T_{\rm CP}^{\rm q}) 
= (329{\rm MeV},46{\rm MeV})$. Thus the change of the critical point
is negligibly small comparing to the hadronic scale.

\subsection{\label{subsec:sus_mu}%
Susceptibilities at finite $\mu_{\rm q}$}
We have previously calculated the quark number density and
susceptibility in the temperature direction. We think it is also
interesting to see the finite quark chemical potential behavior.

\begin{figure}[h!]
\begin{center}
   \includegraphics[width=7.5cm,keepaspectratio]{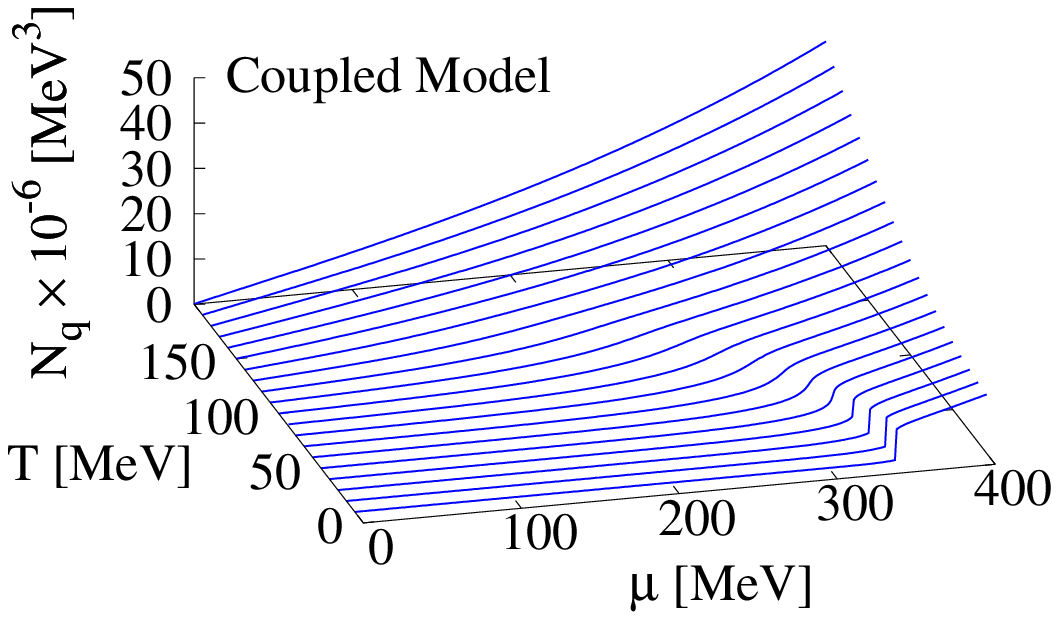}
   \includegraphics[width=7.5cm,keepaspectratio]{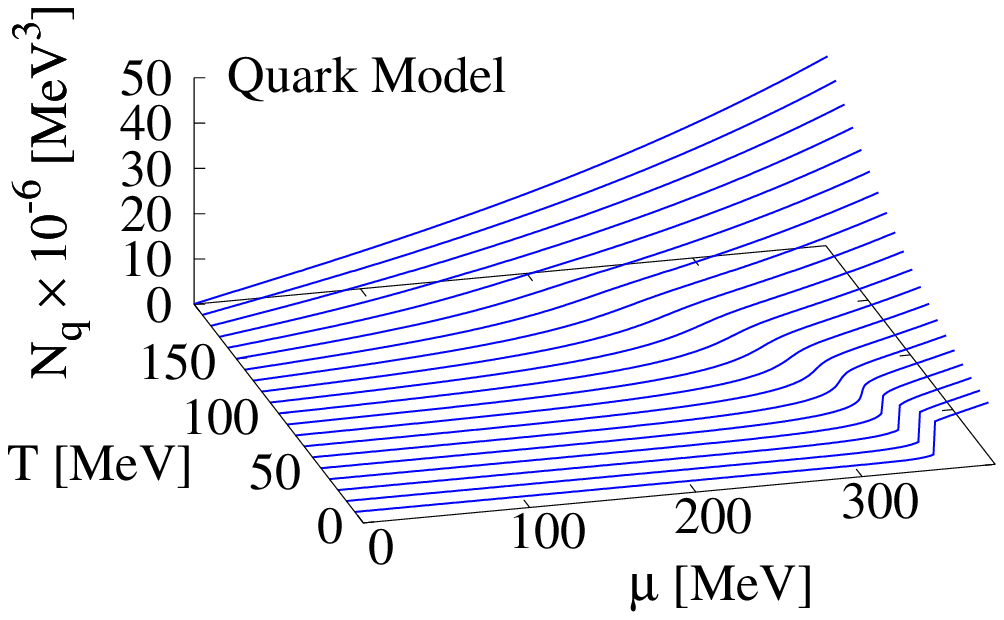}
   \caption{\label{fig:number_3D}
   Quark number density.
   }
\end{center}
\end{figure}

The comparison of the number density on the $\mu$--$T$ plane
within two models is shown in Fig.~\ref{fig:number_3D}.
Nevertheless the deviation at finite $T$ and zero $\mq$ exists
as observed in Figs. \ref{fig:number} and \ref{fig:sus},
the difference is not visually confirmed from the figure; then
we may be able to conclude that the influence from the gluon
condensate is not drastic in two models.

\begin{figure}[h!]
\begin{center}
   \includegraphics[width=7.5cm,keepaspectratio]{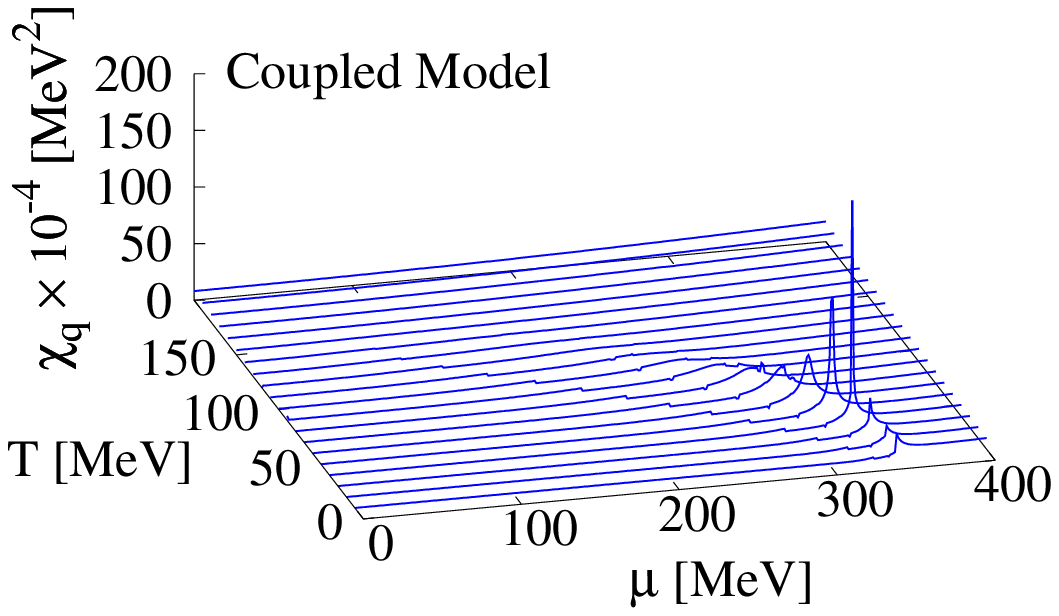}
   \includegraphics[width=7.5cm,keepaspectratio]{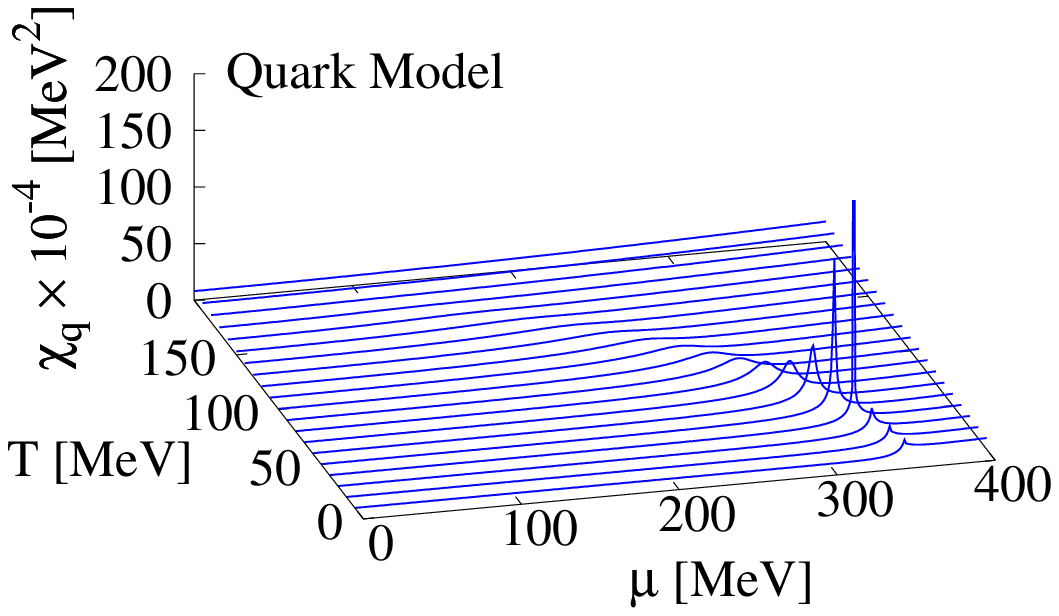}
   \caption{\label{fig:susxeptibility_3D}
   Quark number susceptibility.
   }
\end{center}
\end{figure}

Also, the susceptibility is not considerably different as easily
expected by the above results on the number density, which
is shown in Fig.~\ref{fig:susxeptibility_3D}.

\section{\label{sec:conclusion}%
Conclusions}
We construct the effective model for the gluon and chiral
condensates starting from QCD with the assumed gluon
condensate in this paper. It is found that one can consider
the gap equation of the gluon condensate through evaluating
the effective potential. The solution shows desired behavior
to some extent for low temperature region, while the transition
near the critical temperature is more drastic compare to the
experimental observation of the BEC state. This apparent
discrepancy comes from the crucial distribution change
at high temperature in the Bose-Einstein statistics for bosons
which affects badly to the system of the gap equation. Therefore, 
we believe that the further considerations on the treatment of the
gap equation for the gluon condensate, especially at high
temperature near the transition temperature, should be given so
that the model leads the better description on the phenomena
of the phase transition.

Through the investigation of the coupled model, we studied
how the gluon condensate affects to the chiral sector, particularly
on the thermodynamic quantities and the critical phenomena.
We find that the gluon condensate lowers the transition temperature,
and suppress the quark excitations, i.e., the pressure, number density
and susceptibility. Moreover, the gluon condensate effect reduces the
critical temperature at high quark density. Here, it should be very
important to note that the transition temperature and the location
of the critical point are highly depends on the chosen regularization
produce and parameters as already searched in a lot of preceding
works (see, e.g.,~\cite{Inagaki:2015lma}). Therefore it is necessary to
carefully see the parameter dependence on the phase transition,
which may be the future direction on this model analyses. On the
other hand, the tendency that the gluon condensate suppresses the
quark excitations for low temperature may be the universal feature
of this kind of model with gluon condensate since the suppression
is also seen in the PNJL model studies~\cite{Fukushima:2008wg}.

We think that the current model is still at the primitive level, since
our purpose of the paper is to construct the model incorporating
the gluon condensate based on QCD, then study the model behavior
using the simplest form as the first step analysis. The crucial
point of our simple model is the choice of the effective four coupling,
Eq.~(\ref{eq:replace_co}), in which we set the coupling to be proportional
to the gluon condensate. The numerical results shown in this paper
are mainly come from the choice of the coupling. It may be needed
to consider the form of the coupling as the function of the gluon
condensate, e.g., $G=G(\phi_{\rm g})$,  for the sake of constructing
more realistic model. One more important point in our treatment is
that we do not study the effect from the gluon chemical potential
$\mu_{\rm g}$. In the real situation of the phase transition, the
gluon chemical potential and quark chemical potential relate each
other, then they should be treated simultaneously. This may also
be future direction on this kind of approaches in considering the
gluon and quark condensates.

The above mentioned future generalizations may be required and
important. We believe that, since the current model may have the
possibility of connecting effective models and the first principle
theory of QCD, the further investigations based on this analysis
are interesting.

\begin{acknowledgments}
The author thanks to T. Inagaki and D. Kimura for discussions.
The author is supported by Ministry of Science and Technology
(Taiwan, ROC), through Grant No. MOST 103-2811-M-002-087.
\end{acknowledgments}

\appendix
\section{\label{app:pot}%
The effective potential}
The finite temperature extension of the one-loop contribution
in the effective potential can be performed as follows.

In the imaginary-time formalism, we discretize the
$p_0$-integral,
\begin{align}
       \int \frac{\md^4 p_{E}}{(2\pi)^4} F(p_{E0},{\bf p})
       \to 
       T \sum_{n=-\infty}^{\infty}
       \int \frac{\md^3 p}{(2\pi)^3} F(\omega_n,{\bf p}),
\end{align}
then we need to take the summation 
\begin{align}
       {\mathcal F}
       = T \sum_{n=-\infty}^{\infty}
       \ln \left[  \omega_n^2 + E^2 \right],
\end{align}
with $E^2 = p^2 + M^2$ and $\omega_n = 2 n \pi T$ for bosons
and $\omega_n = (2n+1) \pi T$ for fermions. We first differentiate
the term with respect to $E$ then divide by $2E$,
\begin{align}
       \frac{1}{2E} \frac{d {\mathcal F}}{d E}
       = T \sum_{n=-\infty}^{\infty}
       \frac{1}{\omega_n^2 + E^2},
\end{align}
here we can take the summation as
\begin{align}
       T \sum_{n=-\infty}^{\infty}
       \frac{1}{\omega_n^2 + E^2}
       = \frac{1}{2E}
       \left[ 1 - \frac{1}{e^{\beta E} - s} \right].  
\end{align}
where $s=1$ and $-1$ for the bosons and fermions, respectively.
Thereafter the integration of $d{\mathcal F}/dE$ on $E$ gives
\begin{align}
       {\mathcal F}
       = 
       E   +  T \ln [{1 -s e^{-\beta E}}],
\end{align}
which leads Eqs. (\ref{eq:g_pot_t}) and (\ref{eq:q_pot_t}).

\section{\label{app:sign}%
Sign of the condensates}
We see that the gluon condensate becomes negative if we apply
the three- or four-momentum cutoff method, which
lead the negative effective mass square. This also happens in the
scalar $\phi^4$ theory, and we are going to discuss the
issue of the sign using the $\phi^4$ theory for simplicity.

Let's consider the following Lagrangian,
\begin{align}
       {\mathcal L}_{\phi^4} = 
       \frac{1}{2} (\partial_\mu \phi)^2
       - \frac{1}{4}\lambda \phi^4,
\end{align}
where $\phi$ is the scalar field and $\lambda$ is the four-point
coupling. This gives the effective potential,
\begin{align}
  {\Omega}_{\phi^4}
  = \frac{1}{4}\lambda \phi^4
      -\frac{i}{2} \int \frac{\md^4 p}{(2\pi)^4}
       \ln \left[  -p^2 + M^2 \right],
\end{align}
with $M^2=3\lambda\phi^2$,
then the gap equation ($\partial \Omega/\partial \phi=0$)
leads 
\begin{align}
  \phi^2
  = -3 \int \frac{\md^4 p}{(2\pi)^4}
     \frac{i}{p^2-M^2}.
\label{eq:gap_p4}
\end{align}
It is noted that $\phi^2$ becomes negative if one introduces the
four-momentum cutoff scale $\Lambda$ as,
\begin{align}
  \phi^2
  = -\frac{3}{16\pi^2}
    \left[
      \Lambda^2
      - M^2 \ln \left( \frac{\Lambda^2 + M^2}{M^2} \right)
    \right]
\end{align}
which leads the negative mass square for $\Lambda^2 \gg M^2$.
To remedy the situation, we usually perform the renormalization 
through choosing a specific prescription, e.g., by using the
$\overline{MS}$ scheme we have
\begin{align}
  \phi_{\rm r}^2
  = -\frac{3}{16\pi^2}
    \left[
     M^2 \ln \frac{M^2}{\mu_{\rm r}^2}
    \right],
\end{align}
with the renormalization scale $\mu_{\rm r}$. Note that for
$\mu_{\rm r}^2 > M^2$, $\phi_{\rm r}^2$ becomes positive,
then we can avoid the unwanted negative mass square.

It is worth mentioning that we can have the positive $\Mg^2$
as well for the gluon condensate at $T=0$ if we apply the
renormalization prescription.  However, the system
behaves badly at finite temperature when one considers the
above mentioned renormalization. Here it is useful to employ
the dimensional regularization, because it gives nice description
both at zero and finite temperature contributions simultaneously
thanks to applied analytic continuation.
Therefore, for the sake of treating all the contributions under
consistent way in our effective model analyses, we employ the
dimensional regularization procedure for the gluon condensate.

On the other hand, we have for the fermion field,
\begin{align}
 \phi_{\rm q}=
     -{\rm tr} \int \! \frac{\md^4 q}{(2\pi)^4}
          \frac{i}{q\!\!\!/ - M_{\rm q}}
\end{align}
with $M_{\rm q} = m_{\rm q} -4G\phi_{\rm q}$. Therefore one
needs negative $\phi_{\rm q}$ in order to get the positive effective
mass, and the three- or four-momentum cutoff way nicely works. 
This difference on the sign comes from the fact that
fermion is a Grassmann number, which gives the negative sign
in front of the third line of the effective potential, Eq. (\ref{eq:pot_c}).

\section{\label{app:reg}%
Regularization procedures}
In the current model study,  we employ the two types of 
regularization schemes so as to obtain the positive mass
sign for $\Mg$ and $M_{\rm q}$.

For the gluon integration, as mentioned above, we apply the
dimensional regularization which is defined by
\begin{align}
       \int \frac{\md^4 p}{(2\pi)^4}
       \to 
       M_0^{4-D} \int \frac{\md^D p}{(2\pi)^D},
\end{align}
where $D$ represents the spacetime dimensions, and the
mass scale parameter $M_0$ is introduced to obtain correct
mass dimensions for the integral.

The three-momentum cutoff scheme,
\begin{align}
       \int \frac{\md^4 p}{(2\pi)^4}
       \to 
       \frac{1}{2\pi^2}
       \int_{-\infty}^{\infty} \frac{\md p_0}{2\pi}
       \int_0^{\Lambda} \md p p^2
\end{align}
is the most frequently used method for the quark sector.
We introduce the cutoff scale in the loop integral for zero temperature
contribution, which gives the close curve to the
BCS theory prediction as seen in Fig.~\ref{fig:ratio_q}.


\end{document}